\documentclass[reprint,amsmath,amssymb,aps,prl]{revtex4-1}
\pdfoutput=1
\usepackage{amsmath}
\usepackage{amsfonts}
\usepackage{amssymb}
\usepackage{amsthm}
\usepackage{graphicx}
\usepackage{color}

\newcommand{\ket}[1]{\left\lvert #1 \right\rangle}

\newcommand{\ketbra}[2]{|#1\rangle\langle #2|}
\newcommand{\piover}[1]{{\frac{\pi}{#1}}}

\newcommand{\xor}{\oplus}
\newcommand{\globalxor}[1]{{\oplus_i#1_i}}

\newcommand{\hwp}[1]{U_{HWP}\left(#1\right)}

\newcommand{\COMMENT}[1]{}

\begin{document}

\title{Classical multiparty computation using quantum resources}

\author{Marco Clementi$^{1,2}$, Anna Pappa$^{3,4}$, Andreas Eckstein$^1$, Ian~A.~Walmsley$^1$, Elham Kashefi$^{3,5}$, Stefanie Barz$^{1,6}$}

\affiliation{$^1$~Clarendon Laboratory, Department of Physics, University of Oxford, United Kingdom\\
$^2$~Department of Physics, University of Pavia, Italy\\
$^3$~School of Informatics, University of Edinburgh, United Kingdom\\
$^4$~Department of Physics, University College London, United Kingdom\\
$^5$~LIP6 - CNRS, Universit\'e Pierre et Marie Curie, Paris, France\\
$^6$~ Institute for Functional Matter and Quantum Technologies and Center for Integrated Quantum Science and Technology IQST, University of Stuttgart, Germany}

\date{\today}

\begin{abstract} %300words
In this work, we demonstrate a new way to perform classical multiparty computing amongst parties with limited computational resources. Our method harnesses quantum resources to increase the computational power of the individual parties. We show how a set of clients restricted to linear classical processing are able to jointly compute a non-linear multivariable function that lies beyond their individual capabilities. The clients are only allowed to perform classical XOR gates and single-qubit gates on quantum states. We also examine the type of security that can be achieved in this limited setting. Finally, we provide a proof-of-concept implementation using photonic qubits, that allows four clients to compute a specific example of a multiparty function, the pairwise AND.

\end{abstract}
\maketitle

\section{Introduction}

The ability to communicate and perform computations between parties in a network has become the cornerstone of the modern information age.
As more and more parties with limited resources become connected in wide-scale distributed systems, a critical need is to develop efficient protocols for multiparty computations (MPC), both in terms of communication load and computing capability~\cite{Yao1982,Damgard2006,Bogetoft2009,Saia2015}.

%The focus of this paper is to examine alternate ways 
One approach to efficiently performing MPC is by exploiting quantum resources. It has been shown that  measurements on specific types of entangled states (GHZ states~\cite{Greenberger1989}), when controlled by a linear computer, are sufficient to compute non-linear universal functions~\cite{Anders2009}. Based on that result, it has been demonstrated that MPC under specific assumptions (use of a trusted party, restricted adversaries) is secure, by virtue of the quantum correlations of a GHZ state~\cite{Louko2010}. Similar results have recently been shown in a client-server scenario, where a client restricted to linear (XOR) operations is enabled to securely delegate the computation of a universal boolean function to a quantum server~\cite{Dunjko2014, Barz2016}. The idea behind all these protocols is to use quantum resources in order to compute classical functions more efficiently, without having to build a fully-fledged quantum computer.

%-------------------------------------------------------------------------------------------------------------------
%Figure1 - general scheme
\begin{figure}
	\centering
		\includegraphics[width=0.45\textwidth]{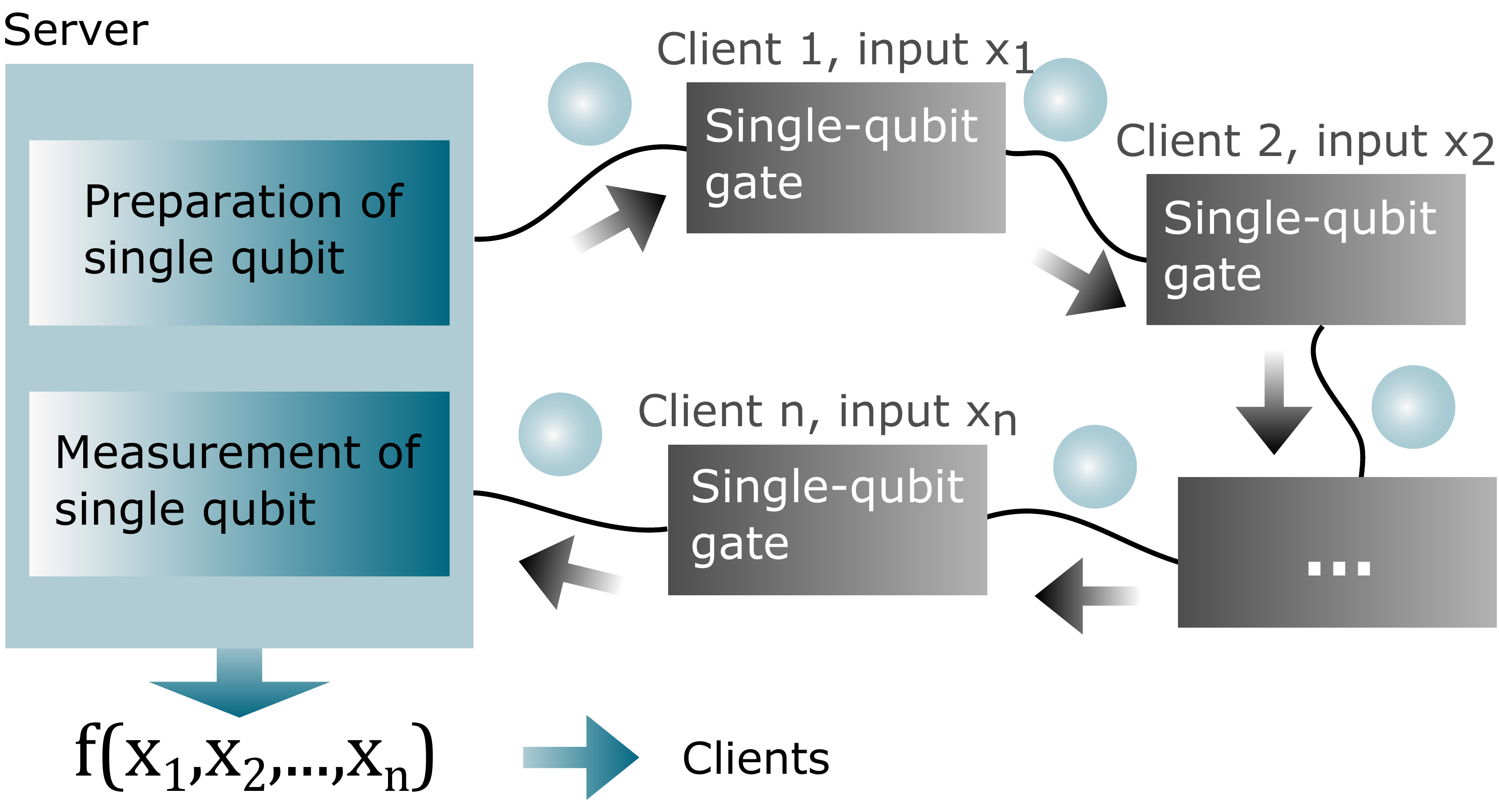}
		\caption{\label{fig:Figure1} A sketch of our scheme for multiparty computation, where a server computes a boolean function $f(x_1, x_2, \ldots,x_n)$ with inputs $x_i$ from different clients. The server generates simple computational resources, such as single qubits, and sends them consecutively to a number of different clients. Each client manipulates the computational resources by performing single qubit gates. At the end, the server measures the output state.
The result of this measurement is sent to the clients, who can deduce the result of the computation.}
\end{figure}
%-------------------------------------------------------------------------------------------------------------------

In this work, we propose a new way of computing non-linear multivariable functions using only linear classical computing and limited manipulation of quantum information.
We examine the scenario where a number of clients want to jointly compute a boolean function of their inputs. We consider that the  clients have limited computing capabilities, namely access to linear (XOR) functionalities. We show how quantum resources can enable such limited clients to securely compute non-linear functions, the complexity of which lies beyond their computing capability. Since access to XOR gates alone is not sufficient for universal classical computing, the clients' computational power is enhanced by means of manipulation of quantum resources provided by a server. 

To demonstrate this boost of computational capabilities using quantum resources, we will focus on a particular example of classical non-linear multiparty computation (the pairwise AND function) that requires as little as one single qubit of communication between the clients. Due to the low quantum communication cost required for the evaluation of this function, the proposed protocol can be used as a building block for more complex computations.

The basic idea of our approach is shown in Fig.~\ref{fig:Figure1}. A quantum server generates a single qubit that is sent through a chain of clients. Each of the clients applies a rotation on the received quantum state according to their classical input. The quantum state is then sent back to the server, which performs a measurement to obtain the result of the computation. Our protocol is designed in such a way that the input of each client remains hidden from the other clients and from the server. Furthermore, the result of the computation remains hidden from the server and is sent back to the clients in an encrypted fashion, meaning that the server performs the computation without learning anything about the result.

%--------------------------------------------------------------------------------------------------------------------------------
%
\section{Theory}
%
%---------------------------------------------------------------------------------------------------------------------------------

Our aim is to compute a non-linear boolean function $f(x_1,\dotsm,x_n)$ on input bits $x_i\in\{0,1\}$. %The clients themselves have limited computing capabilities; they have access to single-qubit gates which perform rotations on single qubits and to local XOR boxes.  
%In this work we will treat the XOR box as a trusted oracle. However, ways of implementing such an oracle under specific security assumptions are present in the literature; for example,~\cite{Louko2010} presents one such case with the parties sharing pairwise randomness. 
We focus on a particular example of a basic multivariable boolean function, the pairwise AND:
\begin{eqnarray}~\label{eqn:function}
f(x_1,\dots,x_n)&=& \bigoplus_{j=1}^n \Big( x_{j+1}\cdot \big(\bigoplus_{i=1}^j x_i\big )\Big )
\end{eqnarray}
%\begin{eqnarray}~\label{eqn:function}
%f(x_1,\dots,x_n)&=&
%x_1\cdot x_2 + (x_1+x_2)\cdot x_3 \nonumber\\
%&&+\dots +(x_1+\dots +x_{n-1})\cdot x_n,
%\end{eqnarray}
The addition and multiplication are the XOR operation and the logical AND operation respectively over the field $\mathbb{F}_2$. If the function in Eqn.~\ref{eqn:function} was linear, then a change in the assignment of one of the variables would either always change the value of the function or would never change it. However it is easy to verify that the function at hand does not follow this rule, and as a non-linear function, it cannot be computed using only linear operations in $\mathbb{F}_2$, such as XOR, but necessitates the use of non-linear operations like NAND. 

Now let us define by $U=R_y(\pi/2)$ the rotation around the $y$-axis of the Bloch sphere (i.e. $R_y(\theta)= e^{-i\theta\sigma_y/2 }$).  Then the following equation can be used to compute the function $f=f(x_1,\dots,x_n)$ in Eqn.~\ref{eqn:function}:
\begin{equation}
\label{wopadding1}
(U^\dagger)^\globalxor{x}U^{x_n}\dotsm U^{x_2}U^{x_1}\ket{0}=\ket{f}
\end{equation}
The fact that Eqn.~\ref{wopadding1} uses only linear processing and operations on a single qubit to compute a non-linear function demonstrates the computational power that quantum resources can provide. Eqn.~\ref{wopadding1} can also be thought of in the clients-server setting described in Fig.~\ref{fig:Figure1}, where each client $C_i$ has an input bit $x_i$ and performs an operation on the received qubit before forwarding it to the next client. By adding extra rotations $V=R_y(\pi)$ around the $y$-axis, we can transform Eqn.~\ref{wopadding1}  in the following way: 
%Here, we aim at computing this function using quantum states and simple single-qubit quantum gates:
%\begin{equation}
%(U^\dagger)^\globalxor{x}U^{x_n}...U^{x_2}U^{x_1}\ket{0}=V^{f(x_1,\dots,x_n)}\ket{0}=\ket{f}
%\label{wopadding}
%\end{equation}, 
%\\
\begin{equation}
\label{wopadding}
(U^\dagger)^\globalxor{x}\\
\underbrace{V^{r_n}U^{x_n}}_{\mathcal{C}_n}\\
...\\
\underbrace{V^{r_2}U^{x_2}}_{\mathcal{C}_2}\\
\underbrace{V^{r_1}U^{x_1}}_{\mathcal{C}_1}\\
\ket{0}=\\
%V^{r}V^{f(x_1,\dots,x_n)}\ket{0}=\\
%V^{r \xor f}\ket{0}=\\
\ket{r \xor f}
\end{equation}
\\
where $r_i$ $\in \{0,1\}$ for $i=1,\dots,n$, and $r=\bigoplus_i r_i$. As we will see in the following sections, this extra $V$ operation will provide some layer of security on top of the computational boost of the clients' power, in the case where there are dishonest participants. %Each of these operations is performed $x_i$ and $r_i$ times respectively by client $C_i$.

%-------------------------------------------------------------------------------------------------------------------
%Figure1 - general scheme
\begin{figure}
	\centering
		\includegraphics[width=0.5\textwidth]{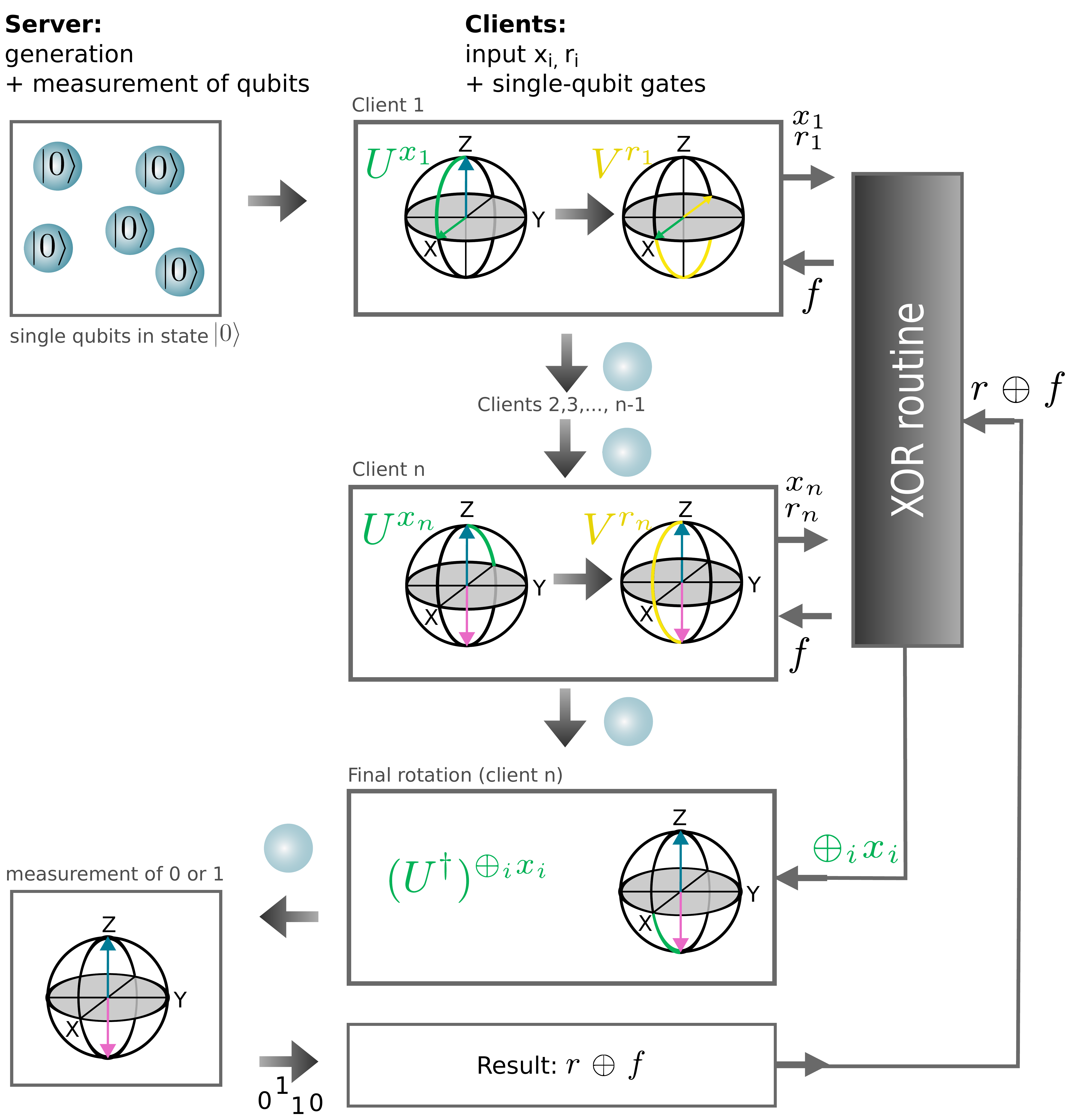}
		\caption{\label{fig:Figure2} Protocol for delegated multiparty computation. For a description of the protocol see main text.}
\end{figure}
%-------------------------------------------------------------------------------------------------------------------

\paragraph{\bf{The protocol.}}
The server generates a single qubit in the state $\ket{0}$ that is sent to client $C_1$.
$C_1$ applies $V^{r_1}U^{x_1}$ on the received qubit, according to input bit $x_1$ and a randomly selected bit $r_1$ and sends the qubit on to the second client $C_2$, who applies $V^{r_2}U^{x_2}$; this procedure continues until all the clients have applied their gates to the qubit (see Figure \ref{fig:Figure2}). The last operation $U^\dagger$ depends on the global XOR of the clients' inputs, computed using a classical routine described below, and can be applied by any client. The resulting state $\ket{r \xor f}$ contains the value of the function up to a random bit flip $r$ (due to Eqn.~\ref{wopadding}).

The qubit is then sent back to the server where a measurement is performed in the computational basis and announces the outcome $r \xor f$. The clients then locally compute the XOR of the random bits of the other clients and perform the last XOR operation $f=r\xor (r\xor f)$ to retrieve the result of the computation.
 
For the computation of the global XOR of both the inputs and the random bits, we consider that the clients run a classical routine that involves using their local XOR boxes to share their classical information among them. During the XOR routine, we assume that the clients communicate between them via secure classical channels that have been established by classical or QKD algorithms.

\paragraph{\bf{The XOR routine.}}
For $i,j=1,\dots,n$, each client $C_j$ uses his local XOR box to choose random bits $x_j^i,~r_j^i\in\{0,1\}$, such that $x_j=\bigoplus_{i=1}^nx_j^i$ and $r_j=\bigoplus_{i=1}^nr_j^i$ and sends $x_j^i$ and $r_j^i$ to client $C_i$. Each client $C_i$ then uses his local XOR box to compute $\tilde{x}_i=\bigoplus_{j=1}^n x_j^i$ and $\tilde{r}_i=\bigoplus_{j=1}^n r_j^i$.
When the designated client needs to perform the operation $U^\dagger$, the rest of the clients send $\tilde{x}_i$ to that client, who uses his local box to compute the global XOR (since $\bigoplus_{i=1}^n x_i=\bigoplus_{i=1}^n \tilde{x}_i$). 

At the end of the protocol, when the server announces the value of the measurement $r\xor f$, all clients broadcast $\tilde{r}_i$, so that all clients can compute the value $r$. Of course, a sequential announcement of the clients will give the last client the ability to learn the output of the computation first, and then abort the protocol. More complicated ways of secret-sharing values and broadcast channels using threshold schemes could be used instead, but that would defeat the purpose of this paper which is to show how simple manipulation of quantum states can boost the computational power of limited clients.

\paragraph{\bf{Security.}}
As mentioned, the goal of this work is to demonstrate how quantum information can increase the computational abilities of parties in a client-server setting; however, the introduction of $V$ rotations in Eqn.~\ref{wopadding} allows for some level of security in a passive adversarial model. More explicitly, we assume that both the server and the clients are interested in completing the computation, and will therefore act according to the protocol; they might however leak some information to an attacker that gains access to their records. We therefore assume that the server sends $\ket{0}$ single-qubit states during the protocol, and no multiple copies of the same qubit or parts of entangled states, but might leak the classical result of the measurement to an eavesdropper. The need to use single copies of quantum states in our protocol is also what distinguishes this work from the previous single-client single-server protocol~\cite{Barz2016}, where using \textit{cobits} (i.e. systems capable of being in a coherent superposition of two states) was sufficient for secure classical computing.

The privacy of the secret input bits of the clients is maintained against someone who acquires a copy of the server's data, since all information that the latter can retrieve is equivalent to the sequence of gates applied, which is in turn equal to $V^{r\xor f}$. Since the term in the exponent represents padded information, the server cannot retrieve more information than that contained in $r\xor f$, which is indeed the expected outcome of measurement.

Furthermore, security against dishonest clients is also maintained, even if we allow them to prepare quantum states and perform measurements on the received states. This is again due to the $V$ rotation that is chosen uniformly at random and performed by all honest clients on the qubit. To see this more clearly, we examine the case when the first honest client in the chain, $C_i$, applies his rotation on the received qubit. We can assume without loss of generality that the qubit is prepared by the dishonest clients in the $XZ$ plane, since all rotations by the honest clients are done on that plane, therefore any component outside the plane will not be affected. The honest operation on any pure state $\ket{\psi}$ on the XZ plane, results in the totally mixed state:
\begin{equation*}
\frac{1}{2}\sum_{r_i}V^{r_i}U^{x_i}\ketbra{\psi}{\psi}(U^\dagger)^{x_i}(V^\dagger)^{r_i}=U^{x_i}\mathbb{I}_2(U^\dagger)^{x_i}
\end{equation*}
which ensures that no information is leaked to the next clients. As in the case of the server however, we need to restrict the clients' malicious behavior to sending single qubit states or equivalently that the honest rotation is done on one qubit. Finally, the client responsible for the last $U^{\dagger}$ rotation will unavoidably learn the parity of the inputs of the rest of the clients, but as long as at least two clients are honest, it is enough to guarantee the secrecy of the independent inputs.

\paragraph{\bf{Efficiency and comparison to previous protocols.}} A common way to perform multiparty computations is via expressing the desired circuits as a sequence of smaller gates, for example 2-input universal gates. Previous work~\cite{Dunjko2014} can therefore be re-interpreted as a protocol that computes the NAND of the inputs of two clients. However, a straightforward extension of this to a multivariable function would prove very costly, requiring one qubit, up to two $R_y$ rotations and several rounds of classical communication to compute the necessary XORs, for each AND evaluation in the function. By just looking at the quantum communication needed in the new protocol (which requires a single qubit to compute the pairwise AND) we observe an immediate gain in efficiency. Furthermore, a straightforward implementation of a construction based on \cite{Dunjko2014} guarantees no security for the inputs of the parties, since the XORs necessary for the application of $U^\dagger$ are on 2 bits, therefore the client who performs the latter unavoidably learns the input of the other client.

Finally, previous studies of boolean function evaluation in the measurement-based quantum computation model~\cite{Hoban2011} required an $(n+1)$-extended GHZ state to compute the pairwise AND function of Eqn. \ref{eqn:function} while to compute other boolean functions (i.e. $n$-tuple AND function), the resource state should have $2^n-1$ qubits. In constrast, the presented protocol does not require any entanglement in the quantum state, and uses only one qubit to compute the pairwise AND function, while for the $n$-tuple AND function, it requires at most $n-1$ qubits (one qubit for each AND operation), giving an exponential decrease on the number of qubits used.

%----------------------------------------------------------------------------------------------------------------------
%Figure 3 - Setup
%----------------------------------------------------------------------------------------------------------------------
\begin{figure}
\begin{center}
\includegraphics[width=0.4\textwidth]{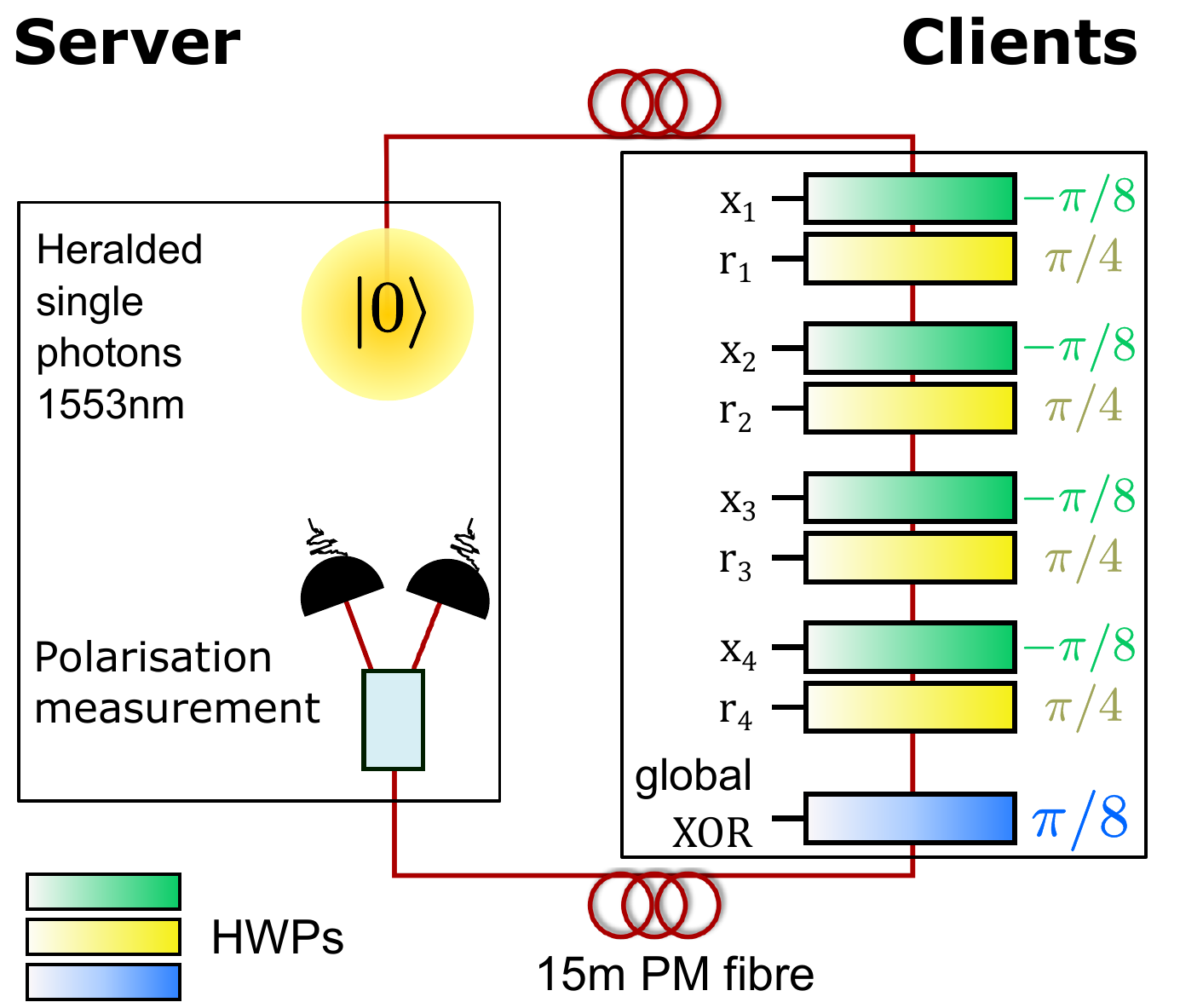}
\end{center}
\caption{Experimental scheme.
The server generates heralded, horizontally polarized photons, which are sent to the clients' side. Each client uses a pair of half-wave plates for applying the gates $V^{r_i}$ and $U^{x_i}$.
If $x_i$ or $r_i$ are equal to zero, the setting of the respective half-wave plate is chosen to be zero. If $x_i$ or $r_i$ are equal to one, the corresponding half-wave plate is rotated by an angle $\theta$ with respect to the horizontal polarisation state, where $\theta$ is given in the Figure.
Finally, one of the clients performs a final conditional rotation dependent on $\globalxor{x}$. The photon is sent back to the server, where a measurement in the computational basis is performed; this has been implemented using a Wollaston prism and two single-photon APDs.}
\label{fig:Figure3}
\end{figure}
%----------------------------------------------------------------------------------------------------------------------

%---------------------------------------------------------------------------------------------------------
\section{Experiment and Results}
%---------------------------------------------------------------------------------------------------------
We implement the protocol using polarisation-encoded photonic qubits with $\ket{0}$ ($\ket{1}$) being the horizontal (vertical) polarisation state. Single photons are generated by pumping a waveguided periodically poled Potassium Titanium Oxide Phosphate crystal with a mode-locked Ti:Sapphire laser ($\tau=200\,$fs, $\lambda=\,$775 nm, 250 kHz repetition rate). After spectral filtering, we obtain pairs of photons at 1547~nm (horizontal polarisation) and 1553~nm (vertical polarisation), each with 2 nm spectral bandwidth (FWHM). The photons are detected using InGaAs avalanche photodetectors (APD)~\cite{Eckstein2011, Harder2013}.

%----------------------------------------------------------------------------------------------------------------------
%Figure 4 - Results
%----------------------------------------------------------------------------------------------------------------------
\begin{figure*}
	\centering
\includegraphics[width=0.9\textwidth]{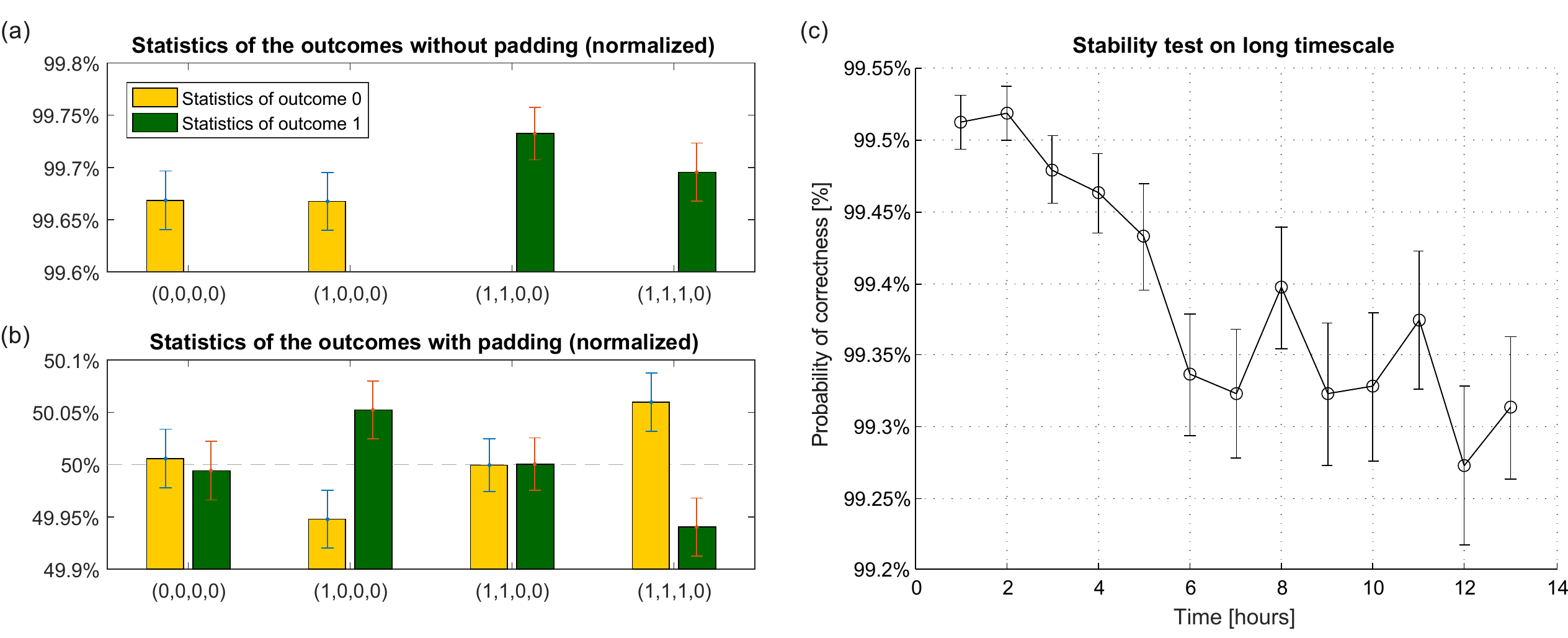}
	\caption{\label{Figure4}(a) Measured  outcomes of the computation after decoding $r\oplus ( r \oplus f )$ for a sample subset $(x_1, x_2, x_3, x_4)$ of the input bits tested  (horizontal axis). (b) Measured outcomes of the computation before decoding $(r\oplus f)$, averaged over all possible combinations of $r_i$, $i=1,\ldots,4$. For each data point, we integrated over 15 seconds, yielding an overall statistics of about 3000 counts for each computation performed. %Error bars are not shown as they are smaller than $0.0003$. 	
	(c) Long term stability of our experiment. The graph shows the probability of obtaining the correct outcome measured over 13 hours of data acquirement. Every point of the plot corresponds to the average over one hour of measurement time. The combination of the clients' input bits used here is $(x_1, x_2, x_3, x_4)$=(1,1,1,1).}
\end{figure*}
%----------------------------------------------------------------------------------------------------------------------

Using this source, the server generates heralded single photons in state $\ket{0}$  which are sent to the clients' side via 15m-long polarisation-maintaining (PM) fibres.
Each client $\mathcal{C}_i$ has access to a series of half-wave plates (HWPs) for implementing the quantum gates $U^{x_i}$ and $V^{r_i}$ (see Fig.~\ref{fig:Figure3}):
\begin{equation}
C_i=\hwp{\piover{4}\cdot r_i} \hwp{-\piover{8}\cdot x_i}
\label{client_hwp}
\end{equation}
where $\hwp{\theta}$ is a HWP with optical axis rotated by $\theta$.  In order to demonstrate all the features of function $f$, we choose to implement a setup with four clients. This could be easily extended straightforwardly to a scheme with an arbitrary number of clients. The overall unitary evolution of the system is then described by the following sequence of operators:

\begin{equation}
\label{client_side_hwp}
\underbrace{\hwp{0}\hwp{\frac{\pi}{8} \cdot \oplus_i x_i}}_{final\,rotation}\\
\underbrace{C_4\,C_3\,C_2\,C_1}_{client\,chain}
\ket{0},
\end{equation}
up to a global phase factor. For the purpose of our demonstration, $\hwp{0}$ can be omitted as it has no effect on the correctness of the demonstration.

Finally, single photons are coupled into another PM fibre and sent back to server. Here, they are measured in the computational basis using a polarisation splitter (extinction ratio $>$60\,dB) and two APDs connected to the output arms.

We performed measurements on all possible 32 sequences of the input bits $x_i$. For each sequence, all possible combinations of the padding bits $r_i$ have been tested. 
Fig.~\ref{Figure4}a shows statistics of the results for a subset of input configurations. 
 The average probability of finding the correct result was measured $(99.53\,\pm\,0.03)\%$, where we assumed Poissonian statistics for the errors. 
 Imperfections arise from state preparation, polarisation manipulation and polarisation measurements, and darks counts.
 Fig.~\ref{Figure4}b shows results for the same input, but averaged over all combinations of random bits $r_i$, resulting in a flat distribution.
The values we obtain for the average outcome of the computation lie between 
$(49.95\,\pm\,0.03)\%$ and $(50.06\,\pm\,0.03)\%$ with an average of $(50.00\,\pm\,0.03)\%$. 
These values are computed from the raw counts corrected by the coupling efficiencies.
This shows that the server could not infer any information from the outcomes of its measurements.
The main limiting factor in the correctness of the result is lies uncertainty in wave-plates positioning and polarisation crosstalk introduced by PM fibre connectors. % ($\pm 8.7$ mrad in our setup). 

Fig.~\ref{Figure4}c shows the long-time stability of our system: we repeated the same computation several times over a time interval of 13 hours and studied drift in our experiment.
The average correctness over this time was $99.43\,\%$ with a standard deviation of $0.08\,\%$.
The correctness decreases from $(99.52\,\pm 0.02)\%$ to $(99.27\pm 0.06)\%$; the drop in probability is caused by drifts in the coupling to the fibres and polarisation drifts.

\paragraph{\bf{Security of implementation.}}
In addition to the theoretical security aspects discussed above, in our implementation we choose the wave-plate settings in such a way that there is no phase shift between the states $\ket{0}$ and $\ket{1}$ that could leak information about the inputs.
 As already discussed in~\cite{Barz2016}, global phase shifts could leak information if the server, for example, sends part of an entangled state.  However, this approach would require an interferometrically stable setup, which is an unlikely condition for a real-life implementation. 
Furthermore, the protocol requires the use of single qubits and a single-shot implementation in order to be secure. For the purpose of computing statistics for our proof-of-principle demonstration, we averaged over several runs of the experiment that used the same input settings. We note, however, that this would leak information about the inputs or the result to a malicious party; therefore in a realistic implementation, single-shot experiments would be required.

%------------------------------------------------------------------------------------------------------------------------------------
\section{Conclusion}
%------------------------------------------------------------------------------------------------------------------------------------
In this work, we demonstrate a novel way to perform non-linear classical multiparty computations by exploiting single qubits and access to restricted linear processes. This is done through studying a specific boolean function that can be thought of as a building block for more complex computations.

Even though the main focus of this work is the boosting of the computational capabilities of limited clients manipulating single qubits, by introducing some extra rotations we can guarantee security under assumptions on the adversarial behavior of the participants. In this setting, the classical data obtained during the protocol do not leak any information, given that the adversaries act in a restricted way. Since the goal was to keep the clients' quantum capabilities as limited as possible, it would defeat the purpose of this study to allow them to perform any check on the correct behavior of the server or the other clients. If we would consider a setting where the clients are enhanced with quantum measurement devices, security of the protocol could be increased by checking the mean photon number (however see \cite{Sajeed2015} for a discussion on attacks and countermeasures on commercial devices).

Our work also offers many avenues for further research. For example, are there more simple non-linear functions like the one presented here that can be used as subroutines for larger computation protocols? And more generally, what is the most efficient way to perform complex computations when we have access to limited quantum and classical resources? Finally, surprisingly enough, this boosting of computational power is possible with the use of single qubits, and without the need of the type of contextuality mentioned in~\cite{Raussendorf2013}, opening a discussion on whether some other form of contextuality is relevant in this setting.


%------------------------------------------------------------------------------------------------------------------------------------
\section{Acknowledgements}
The authors would like to thank Shane Mansfield for very useful suggestions during the development of the theory and Alex E. Jones for comments on the paper.
M.C. acknowledges support from the Erasmus+ programme; A.P. from EPSRC grant EP/M013243/1 and from the European Union's Horizon 2020 Research and Innovation program under Marie Sklodowska-Curie Grant Agreement No. 705194.
I.A.W. acknowledges an ERC Advanced Grant (MOQUACINO) and the UK EPSRC project EP/K034480/1. 
E.K. acknowledges funding through EPSRC funds EP/N003829/1 and EP/M013243/1. 
S.B. acknowledges support from the Marie Curie Actions within the Seventh Framework Programme for Research of the European Commission, under the Initial Training Network PICQUE (Photonic Integrated Compound Quantum Encoding, grant agreement no. 608062) and from the European Union's Horizon 2020 Research and Innovation program under Marie Sklodowska-Curie Grant
Agreement No. 658073.


\begin{thebibliography}{14}%
\makeatletter
\providecommand \@ifxundefined [1]{%
 \@ifx{#1\undefined}
}%
\providecommand \@ifnum [1]{%
 \ifnum #1\expandafter \@firstoftwo
 \else \expandafter \@secondoftwo
 \fi
}%
\providecommand \@ifx [1]{%
 \ifx #1\expandafter \@firstoftwo
 \else \expandafter \@secondoftwo
 \fi
}%
\providecommand \natexlab [1]{#1}%
\providecommand \enquote  [1]{``#1''}%
\providecommand \bibnamefont  [1]{#1}%
\providecommand \bibfnamefont [1]{#1}%
\providecommand \citenamefont [1]{#1}%
\providecommand \href@noop [0]{\@secondoftwo}%
\providecommand \href [0]{\begingroup \@sanitize@url \@href}%
\providecommand \@href[1]{\@@startlink{#1}\@@href}%
\providecommand \@@href[1]{\endgroup#1\@@endlink}%
\providecommand \@sanitize@url [0]{\catcode `\\12\catcode `\$12\catcode
  `\&12\catcode `\#12\catcode `\^12\catcode `\_12\catcode `\%12\relax}%
\providecommand \@@startlink[1]{}%
\providecommand \@@endlink[0]{}%
\providecommand \url  [0]{\begingroup\@sanitize@url \@url }%
\providecommand \@url [1]{\endgroup\@href {#1}{\urlprefix }}%
\providecommand \urlprefix  [0]{URL }%
\providecommand \Eprint [0]{\href }%
\providecommand \doibase [0]{http://dx.doi.org/}%
\providecommand \selectlanguage [0]{\@gobble}%
\providecommand \bibinfo  [0]{\@secondoftwo}%
\providecommand \bibfield  [0]{\@secondoftwo}%
\providecommand \translation [1]{[#1]}%
\providecommand \BibitemOpen [0]{}%
\providecommand \bibitemStop [0]{}%
\providecommand \bibitemNoStop [0]{.\EOS\space}%
\providecommand \EOS [0]{\spacefactor3000\relax}%
\providecommand \BibitemShut  [1]{\csname bibitem#1\endcsname}%
\let\auto@bib@innerbib\@empty
%</preamble>
\bibitem [{\citenamefont {Yao}(1982)}]{Yao1982}%
  \BibitemOpen
  \bibfield  {author} {\bibinfo {author} {\bibfnamefont {A.~C.}\ \bibnamefont
  {Yao}},\ }in\ \href@noop {} {\emph {\bibinfo {booktitle} {Proceedings of the
  23rd Annual Symposium on Foundations of Computer Science}}},\ \bibinfo
  {series and number} {SFCS '82}\ (\bibinfo  {publisher} {IEEE Computer
  Society},\ \bibinfo {year} {1982})\ pp.\ \bibinfo {pages}
  {160--164}\BibitemShut {NoStop}%
\bibitem [{\citenamefont {Damg{\aa}rd}(1982)}]{Damgard2006}%
  \BibitemOpen
  \bibfield  {author} {\bibinfo {author} {\bibfnamefont {I.}~\bibnamefont
  {Damg{\aa}rd}},\ }in\ \href@noop {} {\emph {\bibinfo {booktitle} {Proceedings
  of the 5th International Conference on Security and Cryptography for
  Networks}}},\ \bibinfo {series and number} {SFCS '82}\ (\bibinfo  {publisher}
  {Springer Berlin Heidelberg},\ \bibinfo {year} {1982})\ pp.\ \bibinfo {pages}
  {360--364}\BibitemShut {NoStop}%
\bibitem [{\citenamefont {Bogetoft}\ \emph {et~al.}(2009)\citenamefont
  {Bogetoft}, \citenamefont {Christensen}, \citenamefont {Damg{\aa}rd},
  \citenamefont {Geisler}, \citenamefont {Jakobsen}, \citenamefont
  {Kr{\o}igaard}, \citenamefont {Nielsen}, \citenamefont {Nielsen},
  \citenamefont {Nielsen}, \citenamefont {Pagter}, \citenamefont
  {Schwartzbach},\ and\ \citenamefont {Toft}}]{Bogetoft2009}%
  \BibitemOpen
  \bibfield  {author} {\bibinfo {author} {\bibfnamefont {P.}~\bibnamefont
  {Bogetoft}}, \bibinfo {author} {\bibfnamefont {D.~L.}\ \bibnamefont
  {Christensen}}, \bibinfo {author} {\bibfnamefont {I.}~\bibnamefont
  {Damg{\aa}rd}}, \bibinfo {author} {\bibfnamefont {M.}~\bibnamefont
  {Geisler}}, \bibinfo {author} {\bibfnamefont {T.}~\bibnamefont {Jakobsen}},
  \bibinfo {author} {\bibfnamefont {M.}~\bibnamefont {Kr{\o}igaard}}, \bibinfo
  {author} {\bibfnamefont {J.~D.}\ \bibnamefont {Nielsen}}, \bibinfo {author}
  {\bibfnamefont {J.~B.}\ \bibnamefont {Nielsen}}, \bibinfo {author}
  {\bibfnamefont {K.}~\bibnamefont {Nielsen}}, \bibinfo {author} {\bibfnamefont
  {J.}~\bibnamefont {Pagter}}, \bibinfo {author} {\bibfnamefont
  {M.}~\bibnamefont {Schwartzbach}}, \ and\ \bibinfo {author} {\bibfnamefont
  {T.}~\bibnamefont {Toft}},\ }in\ \href@noop {} {\emph {\bibinfo {booktitle}
  {Proceedings of the 13th International Conference on Financial Cryptography
  and Data Security}}},\ \bibinfo {series and number} {FC '09}\ (\bibinfo
  {publisher} {Springer Berlin Heidelberg},\ \bibinfo {year} {2009})\ pp.\
  \bibinfo {pages} {325--343}\BibitemShut {NoStop}%
\bibitem [{\citenamefont {Saia}\ and\ \citenamefont {Zamani}(2015)}]{Saia2015}%
  \BibitemOpen
  \bibfield  {author} {\bibinfo {author} {\bibfnamefont {J.}~\bibnamefont
  {Saia}}\ and\ \bibinfo {author} {\bibfnamefont {M.}~\bibnamefont {Zamani}},\
  }in\ \href@noop {} {\emph {\bibinfo {booktitle} {Proceedings of the 41st
  International Conference on Current Trends in Theory and Practice of Computer
  Science}}}\ (\bibinfo  {publisher} {Springer Berlin Heidelberg},\ \bibinfo
  {year} {2015})\ pp.\ \bibinfo {pages} {24--44}\BibitemShut {NoStop}%
\bibitem [{\citenamefont {Greenberger}\ \emph {et~al.}(1989)\citenamefont
  {Greenberger}, \citenamefont {Horne},\ and\ \citenamefont
  {Zeilinger}}]{Greenberger1989}%
  \BibitemOpen
  \bibfield  {author} {\bibinfo {author} {\bibfnamefont {D.~M.}\ \bibnamefont
  {Greenberger}}, \bibinfo {author} {\bibfnamefont {M.~A.}\ \bibnamefont
  {Horne}}, \ and\ \bibinfo {author} {\bibfnamefont {A.}~\bibnamefont
  {Zeilinger}},\ }\enquote {\bibinfo {title} {Going beyond bell's theorem},}\
  in\ \href@noop {} {\emph {\bibinfo {booktitle} {Bell's Theorem, Quantum
  Theory, and Conceptions of the Universe}}},\ \bibinfo {editor} {edited by\
  \bibinfo {editor} {\bibfnamefont {M.}~\bibnamefont {Kafatos}}}\ (\bibinfo
  {publisher} {Kluwer},\ \bibinfo {address} {Dordrecht},\ \bibinfo {year}
  {1989})\ pp.\ \bibinfo {pages} {73--76}\BibitemShut {NoStop}%
\bibitem [{\citenamefont {Anders}\ and\ \citenamefont
  {Browne}(2009)}]{Anders2009}%
  \BibitemOpen
  \bibfield  {author} {\bibinfo {author} {\bibfnamefont {J.}~\bibnamefont
  {Anders}}\ and\ \bibinfo {author} {\bibfnamefont {D.~E.}\ \bibnamefont
  {Browne}},\ }\href@noop {} {\bibfield  {journal} {\bibinfo  {journal}
  {Physical Review Letters}\ }\textbf {\bibinfo {volume} {102}},\ \bibinfo
  {pages} {050502} (\bibinfo {year} {2009})}\BibitemShut {NoStop}%
\bibitem [{\citenamefont {Loukopoulos}\ and\ \citenamefont
  {Browne}(2010)}]{Louko2010}%
  \BibitemOpen
  \bibfield  {author} {\bibinfo {author} {\bibfnamefont {K.}~\bibnamefont
  {Loukopoulos}}\ and\ \bibinfo {author} {\bibfnamefont {D.~E.}\ \bibnamefont
  {Browne}},\ }\href {\doibase 10.1103/PhysRevA.81.062336} {\bibfield
  {journal} {\bibinfo  {journal} {Phys. Rev. A}\ }\textbf {\bibinfo {volume}
  {81}},\ \bibinfo {pages} {062336} (\bibinfo {year} {2010})}\BibitemShut
  {NoStop}%
\bibitem [{\citenamefont {Dunjko}\ \emph {et~al.}(2016)\citenamefont {Dunjko},
  \citenamefont {Kapourniotis},\ and\ \citenamefont {Kashefi}}]{Dunjko2014}%
  \BibitemOpen
  \bibfield  {author} {\bibinfo {author} {\bibfnamefont {V.}~\bibnamefont
  {Dunjko}}, \bibinfo {author} {\bibfnamefont {T.}~\bibnamefont
  {Kapourniotis}}, \ and\ \bibinfo {author} {\bibfnamefont {E.}~\bibnamefont
  {Kashefi}},\ }\href@noop {} {\bibfield  {journal} {\bibinfo  {journal}
  {Journal of Quantum Information and Computation}\ ,\ \bibinfo {pages} {0061}}
  (\bibinfo {year} {2016})}\BibitemShut {NoStop}%
\bibitem [{\citenamefont {Barz}\ \emph {et~al.}(2016)\citenamefont {Barz},
  \citenamefont {Dunjko}, \citenamefont {Schlederer}, \citenamefont {Moore},
  \citenamefont {Kashefi},\ and\ \citenamefont {Walmsley}}]{Barz2016}%
  \BibitemOpen
  \bibfield  {author} {\bibinfo {author} {\bibfnamefont {S.}~\bibnamefont
  {Barz}}, \bibinfo {author} {\bibfnamefont {V.}~\bibnamefont {Dunjko}},
  \bibinfo {author} {\bibfnamefont {F.}~\bibnamefont {Schlederer}}, \bibinfo
  {author} {\bibfnamefont {M.}~\bibnamefont {Moore}}, \bibinfo {author}
  {\bibfnamefont {E.}~\bibnamefont {Kashefi}}, \ and\ \bibinfo {author}
  {\bibfnamefont {I.~A.}\ \bibnamefont {Walmsley}},\ }\href@noop {} {\bibfield
  {journal} {\bibinfo  {journal} {Physical Review A}\ }\textbf {\bibinfo
  {volume} {93}},\ \bibinfo {pages} {032339} (\bibinfo {year}
  {2016})}\BibitemShut {NoStop}%
\bibitem [{\citenamefont {Hoban}\ \emph {et~al.}(2011)\citenamefont {Hoban},
  \citenamefont {Campbell}, \citenamefont {Loukopoulos},\ and\ \citenamefont
  {Browne}}]{Hoban2011}%
  \BibitemOpen
  \bibfield  {author} {\bibinfo {author} {\bibfnamefont {M.~J.}\ \bibnamefont
  {Hoban}}, \bibinfo {author} {\bibfnamefont {E.~T.}\ \bibnamefont {Campbell}},
  \bibinfo {author} {\bibfnamefont {K.}~\bibnamefont {Loukopoulos}}, \ and\
  \bibinfo {author} {\bibfnamefont {D.~E.}\ \bibnamefont {Browne}},\
  }\href@noop {} {\bibfield  {journal} {\bibinfo  {journal} {New Journal of
  Physics}\ }\textbf {\bibinfo {volume} {13}},\ \bibinfo {pages} {023014}
  (\bibinfo {year} {2011})}\BibitemShut {NoStop}%
\bibitem [{\citenamefont {Eckstein}\ \emph {et~al.}(2011)\citenamefont
  {Eckstein}, \citenamefont {Christ}, \citenamefont {Mosley},\ and\
  \citenamefont {Silberhorn}}]{Eckstein2011}%
  \BibitemOpen
  \bibfield  {author} {\bibinfo {author} {\bibfnamefont {A.}~\bibnamefont
  {Eckstein}}, \bibinfo {author} {\bibfnamefont {A.}~\bibnamefont {Christ}},
  \bibinfo {author} {\bibfnamefont {P.~J.}\ \bibnamefont {Mosley}}, \ and\
  \bibinfo {author} {\bibfnamefont {C.}~\bibnamefont {Silberhorn}},\ }\href
  {\doibase 10.1103/PhysRevLett.106.013603} {\bibfield  {journal} {\bibinfo
  {journal} {Phys. Rev. Lett.}\ }\textbf {\bibinfo {volume} {106}},\ \bibinfo
  {pages} {013603} (\bibinfo {year} {2011})}\BibitemShut {NoStop}%
\bibitem [{\citenamefont {Harder}\ \emph {et~al.}(2013)\citenamefont {Harder},
  \citenamefont {Ansari}, \citenamefont {Brecht}, \citenamefont {Dirmeier},
  \citenamefont {Marquardt},\ and\ \citenamefont {Silberhorn}}]{Harder2013}%
  \BibitemOpen
  \bibfield  {author} {\bibinfo {author} {\bibfnamefont {G.}~\bibnamefont
  {Harder}}, \bibinfo {author} {\bibfnamefont {V.}~\bibnamefont {Ansari}},
  \bibinfo {author} {\bibfnamefont {B.}~\bibnamefont {Brecht}}, \bibinfo
  {author} {\bibfnamefont {T.}~\bibnamefont {Dirmeier}}, \bibinfo {author}
  {\bibfnamefont {C.}~\bibnamefont {Marquardt}}, \ and\ \bibinfo {author}
  {\bibfnamefont {C.}~\bibnamefont {Silberhorn}},\ }\href@noop {} {\bibfield
  {journal} {\bibinfo  {journal} {Optics express}\ }\textbf {\bibinfo {volume}
  {21}},\ \bibinfo {pages} {13975} (\bibinfo {year} {2013})}\BibitemShut
  {NoStop}%
\bibitem [{\citenamefont {Sajeed}\ \emph {et~al.}(2015)\citenamefont {Sajeed},
  \citenamefont {Radchenko}, \citenamefont {Kaiser}, \citenamefont {Bourgoin},
  \citenamefont {Pappa}, \citenamefont {Monat}, \citenamefont {Legr\'e},\ and\
  \citenamefont {Makarov}}]{Sajeed2015}%
  \BibitemOpen
  \bibfield  {author} {\bibinfo {author} {\bibfnamefont {S.}~\bibnamefont
  {Sajeed}}, \bibinfo {author} {\bibfnamefont {I.}~\bibnamefont {Radchenko}},
  \bibinfo {author} {\bibfnamefont {S.}~\bibnamefont {Kaiser}}, \bibinfo
  {author} {\bibfnamefont {J.-P.}\ \bibnamefont {Bourgoin}}, \bibinfo {author}
  {\bibfnamefont {A.}~\bibnamefont {Pappa}}, \bibinfo {author} {\bibfnamefont
  {L.}~\bibnamefont {Monat}}, \bibinfo {author} {\bibfnamefont
  {M.}~\bibnamefont {Legr\'e}}, \ and\ \bibinfo {author} {\bibfnamefont
  {V.}~\bibnamefont {Makarov}},\ }\href {\doibase 10.1103/PhysRevA.91.032326}
  {\bibfield  {journal} {\bibinfo  {journal} {Phys. Rev. A}\ }\textbf {\bibinfo
  {volume} {91}},\ \bibinfo {pages} {032326} (\bibinfo {year}
  {2015})}\BibitemShut {NoStop}%
\bibitem [{\citenamefont {Raussendorf}(2013)}]{Raussendorf2013}%
  \BibitemOpen
  \bibfield  {author} {\bibinfo {author} {\bibfnamefont {R.}~\bibnamefont
  {Raussendorf}},\ }\href {\doibase 10.1103/PhysRevA.88.022322} {\bibfield
  {journal} {\bibinfo  {journal} {Phys. Rev. A}\ }\textbf {\bibinfo {volume}
  {88}},\ \bibinfo {pages} {022322} (\bibinfo {year} {2013})}\BibitemShut
  {NoStop}%
\end{thebibliography}
\end{document}